\documentclass[aps,showpacs,letterpaper,twocolumn,12point]{revtex4-1}

\usepackage{graphicx} 
\usepackage{amsmath}
\usepackage{color}      

\usepackage{amstext}
\usepackage{graphics}
\usepackage{amssymb}

\newcommand \be{\begin{equation}}
\newcommand \ee{\end{equation}}
\newcommand \bea{\begin{eqnarray}}
\newcommand \eea{\end{eqnarray}}
\newcommand \bse{\begin{subequations}}
\newcommand \ese{\end{subequations}}
\newcommand \nn{\nonumber}

\newcommand \mcE{{\mathcal E}}
\newcommand \mcF{{\mathcal F}}

\begin{document}

\title{Atom trapping in a bottle beam created by a diffractive optical element}

\author{V. V. Ivanov, J. A. Isaacs, M. Saffman} 

\address{
Department of Physics, University of Wisconsin, 1150 University Avenue, Madison, Wisconsin 53706, USA
}
\author{S.A. Kemme, A.R. Ellis, G.R. Brady, J.R. Wendt, G. W. Biedermann, and S. Samora}
\address{
Sandia National Laboratories, Albuquerque, NM  87185-1082, USA
}

\begin{abstract} A diffractive optical element (DOE) has been fabricated for  creating blue detuned atomic bottle beam traps. The DOE integrates several diffractive lenses for trap creation and imaging of atomic fluorescence. We characterize the performance of the DOE and demonstrate trapping of cold Cesium atoms inside a bottle beam. \end{abstract}

\date{\today}

\maketitle

\section{Introduction}

Atomic qubits localized in an array of optical traps represent
a promising and actively pursued approach to implementing multi-qubit quantum 
information processing (QIP) devices\cite{Meschede2006,Saffman2010,Negretti2011,Schlosser2011}.  
Far detuned optical traps provide strong confinement with low photon scattering rates and low decoherence. Optical trap arrays have been synthesized 
with many different optical configurations, most often using bulk optical elements. Replacing bulk optics 
with chip scale components \cite{Streed2011,Kim2011,Brady2011} has several potential advantages, including size, robust alignment, integration with other system components including vacuum chambers, and convenient scalability to many sites. 

Several recent experiments have demonstrated single atom loading  in blue detuned traps created with bulk optics\cite{Xu2010,Li2012}. 
 We are particularly interested in blue detuned traps in which atoms are localized at a minimum of the optical intensity since they enable simultaneous and magic trapping of atoms in ground and Rydberg states\cite{SZhang2011}. This capability will be important for future scalable QIP devices based on Rydberg state mediated quantum gates\cite{Wilk2010,Isenhower2010,Zhang2010} as well as adiabatic approaches based on Rydberg dressing\cite{Keating2013} or dissipative interactions\cite{Carr2013} which require long term occupancy of Rydberg states.

\begin{figure}[!t]
\begin{center}
\includegraphics[width=.45\textwidth]
{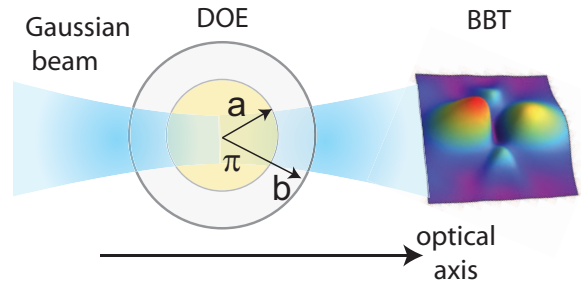}
\caption{(color online) An incident Gaussian beam is transformed into a BBT using a DOE that combines the focusing and phase shifting functions needed to make a BBT. }
\label{fig.sketch}
\end{center}
\end{figure}

We present here a diffractive optical element for creating  a blue detuned bottle beam trap (BBT). Our approach combines the $\pi$ phase plate and focusing lens of \cite{Chaloupka1997,Ozeri1999,*Ozeri2002} into a single diffractive element etched in fused silica. As shown in 
Fig. \ref{fig.sketch} a Gaussian TEM$_{00}$ beam incident on the DOE is transformed into a BBT located a distance $z\simeq f$ after the DOE, where $f$ is the focal 
length of the lens. The trap formation part of the DOE was designed for a wavelength of 780 nm, which is blue detuned from the Cs D lines at 852 and 894 nm. In addition the DOE contains an outer lens designed for a wavelength of 852 nm which can collect fluorescence light from an atom localized in the BBT, which will be useful for qubit detection. The focal point of the collection lens  is accurately aligned with the position of the BBT due to the lithographic precision of the DOE fabrication process.

The rest of this paper is organized as follows. In Sec. \ref{sec.doe} we provide a detailed description of the DOE together with a theoretical account 
of the resulting BBT based on Fresnel diffraction theory. In Sec. \ref{sec.doemeasurements} we present measurements of the BBT intensity distribution and compare with theory. We also characterize the performance of the collection lens. In Sec. \ref{sec.atoms} we demonstrate trapping of Cs atoms, and conclude with a future outlook in Sec. \ref{sec.end}.

\section{DOE for creating a bottle beam trap}

\label{sec.doe}

\subsection{DOE design and fabrication}

\begin{figure}[!t]
\centering
\includegraphics[width=.45\textwidth]
{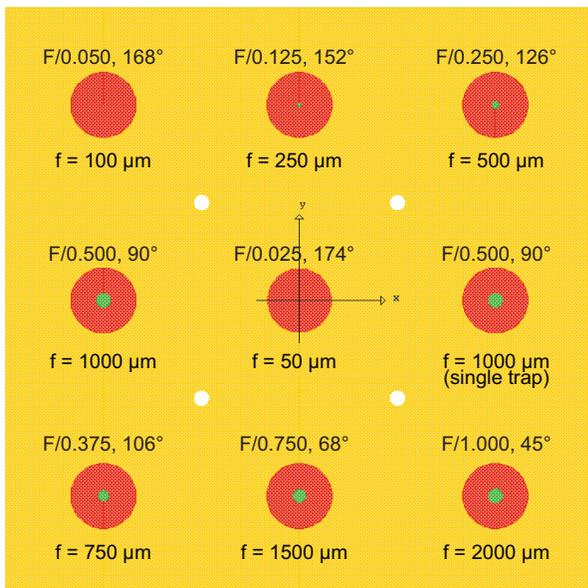}
\caption{(color online) Wafer of 9 BBT and collection diffractive optical elements on a 1" square. Each DOE is labeled with the F/\# plus full cone angle of the surrounding collection optics (above) and the focal length (below).  }
\label{fig.wafer}
\end{figure}

A single die carrying several DOEs for bottle beam trap creation and fluorescence imaging was fabricated on a fused silica substrate\cite{Kemme2012} as shown in Fig. \ref{fig.wafer}. 
Each DOE consists of two regions: an inner region (green in the figure) which creates a bottle beam trap with 
780 nm light and an
outer region (red in the figure) which is a Fresnel lens for 850 nm light with a focal distance identical to the bottle beam lens. Thus  
fluorescence from  atoms inside the BBT will be collected and collimated. Such an 
arrangement combines a large numerical aperture with a compact and robust optical setup. The region in between the lenses is  gold coated for use in forming a mirror magneto-optical trap. 

\begin{figure}[!t]
\centering
\includegraphics[width=.45\textwidth]
{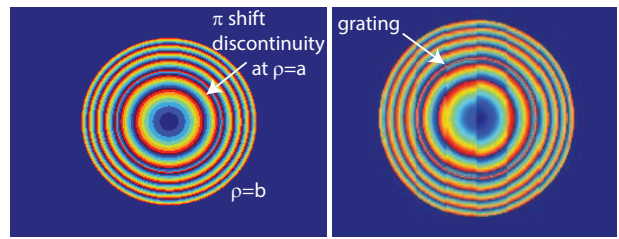}
\caption{Diffractive optical element with 8 levels and a $\pi$ phase shift at the indicated radius $a$ (left) and diffractive lens with a beam splitting grating superimposed (right).   }
\label{fig.doelens}
\end{figure}

Both the bottle beam trapping DOE as well as the fluorescence collection DOE were fabricated monolithically on the same fused silica substrate.  Each 8-level diffractive bottle beam trap DOE is integrated with and functions coaxially with the corresponding ultra-fast collection DOE with the same focal position.  The bottle beam trapping DOE is implemented  as an 8-level diffractive element with a  $\pi$ phase shift radius $a$ as shown in Fig. \ref{fig.doelens} (left).
Illumination of the DOE with a Gaussian beam with waist radius $w_0=a/\sqrt{\ln2}$ creates an on-axis intensity null at the back focal plane of the lens.  In order to implement the phase shift, it is only necessary to radially shift the rings forming the diffractive lens at the appropriate radius $a$. The outer radius $b$ is chosen large enough such that essentially all the energy in the illuminating Gaussian beam passes through the lens.  A double trap can be formed by adding  a linear grating as shown in Fig. \ref{fig.doelens} (right). A grating period of $\Lambda= \lambda/[2\sin(\tan^{-1}(s/2f))]$ produces two traps separated by a distance $s$ in the back focal plane of the lens. 

The 8-level diffractive bottle beam, at a design wavelength of 780 nm, is surrounded by and functions coaxially with an ultra-fast lens of the same focal length but at a design wavelength of 852 nm for collection of fluorescence light from trapped Cs atoms.  The collection lenses are all 2 mm in diameter and are  also realized as  8-level diffractive elements even though all of the collection elements on the wafer are at least as fast as F/1.  The F/1 DOE collects over a full angle of 45 degrees, while the most aggressive DOE on the wafer operates at F/0.025 and collects over slightly less than the hemisphere (172 degrees full angle).   These collection lenses are to our knowledge 
the fastest diffractive elements yet realized, ranging from F/1 down to F/0.025. They are substantially faster than the 
2-level microoptic at F/0.8 used for fluorescence collection from trapped ions\cite{Streed2011} or the 8-level F/0.6 diffractive optical elements integrated into an ion trap\cite{Brady2011}.
The greatest fabrication challenge was in realizing a mask writing/liftoff process that can accommodate the wide range of feature sizes present from the DOE mesas at the edge of the 1 mm radius collection DOEs, all at uniform etch depths across the wafer.

\subsection{Bottle beam DOE analysis}

\begin{figure}[!t]
\begin{center}
\includegraphics[width=.45\textwidth]
{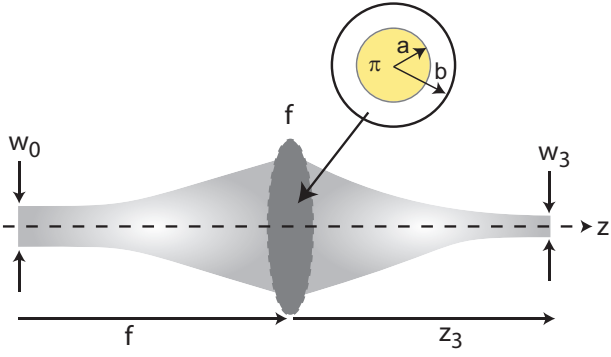}
\caption{(color online) Gaussian beam illuminating a lens with a $\pi$ phase in the inner region of radius $b$. }
\label{fig.DOEBBT}
\end{center}
\end{figure}

We now present a detailed analysis of light diffraction in the bottle beam forming DOE. 
The optical layout 
is shown in Fig. \ref{fig.DOEBBT}. The incident TEM$_{00}$ beam with waist 
$w_0$ a distance $f$ in front of the lens is transformed into a new waist $w_3$ at position $z_3$ after the lens.  For this problem with radial symmetry 
it is most convenient to analyze the diffractive propagation of the optical field $A(\rho)$ using the angular spectrum $\tilde A(\kappa)$ which can be expressed  as a Hankel transform 
\be
\tilde A(\kappa)= \int\limits_{0}^{\infty} d\rho\, \rho A(\rho) J_0(\rho \kappa),
\label{eq.FBessel1}
\ee
with the inverse transform
\be
A(\rho)= \int\limits_{0}^{\infty} d \kappa \, \kappa \tilde A(\kappa) J_0(\rho \kappa).
\label{eq.FBessel2}
\ee
Here $J_0$ is the zeroth order Bessel function, $\rho$ is the radial coordinate, and $\kappa$ is the radial wavenumber.

Call the field in the front focal plane $A_0$ then at the lens the field is 
$$
A_1 = \mcF^{-1}[\mcF[A_0(\rho)] H(\kappa,f)]
$$
with the Fresnel kernel $H(\kappa,z)=e^{-\imath \frac{\kappa^2 z}{2k}}$, $k=2\pi/\lambda$, and $\lambda$ is the wavelength in vacuum.  Passage through the thin  lens with two annular regions is accounted for by multiplying with the transmission function 
$$
t(\rho)=e^{-\imath \frac{k \rho^2 }{2f}}\left[1-2\, {\rm circ}(\rho/a)\right]
$$
where  we have introduced the radial step function ${\rm circ}(\rho)=1$ for $\rho\le 1$ and ${\rm circ}(\rho)=0$ for $\rho> 1.$ 
We will assume the field does not extend to the outer boundary of the lens and ignore the outer radius $b$.
We thus get $A_2(\rho)=t A_1(\rho)$ and the field in the output plane a distance $z$ after the lens is given by 
\bea
&&A_3(\rho_3,z)\nonumber\\
 &=& \mcF_{32}^{-1}[\mcF_{22}[A_2(\rho_2)] H(\kappa_{2},z)]\nn\\
&=&\mcF_{32}^{-1}[\mcF_{22}[t(\rho_1)\mcF_{10}^{-1}[\mcF_{00}[A_0(\rho_0)] H(\kappa_{0},f)]] H(\kappa_{2},z)]\nonumber\\.
\label{eq.bb2}
\eea
Subscripts on the variables and operators have been introduced to indicate different transverse coordinates corresponding to the  planes in Fig. \ref{fig.DOEBBT}. The operator $\mcF_{kj}$ connects $\rho_j$ with $\kappa_{k}$ and $\mcF_{kj}^{-1}$ connects $\kappa_j$ with $\rho_{k}$.
 Writing out (\ref{eq.bb2}) explicitly the output field can be expressed as 
\begin{widetext}
\bea
A_3(\rho_3,z_3) &=&\int d\kappa_{1}\, \kappa_{1} J_0(\rho_3 \kappa_{1}) e^{-\imath \frac{\kappa_{1}^2 z_3}{2k}}
\int d\rho_1\,  \rho_{1} J_0(\rho_1 \kappa_{1}) 
e^{-\imath \frac{k \rho_1^2 }{2f}}\left[1-2\, {\rm circ}(\rho_1/a)\right]\nn\\
&&\times 
\int d\kappa_{0} \kappa_{0} J_0(\rho_1 \kappa_{0})  e^{-\imath \frac{\kappa_{0}^2 f}{2k}}
\int d\rho_0\, \rho_{0} J_0(\rho_0 \kappa_{0})  A_0(\rho_0) 
\label{eq.bb3}
\eea
\end{widetext}

The factor $\left[1-2\, {\rm circ}(\rho_1/a)\right]$ splits the result into two terms. The first one, which is independent of $a$,
results in a Gaussian beam  with waist $w_3= \lambda f/(\pi w_0)$ at $z_3=f$
\be
A_{3b}(\rho_3,z_3)=-i\frac{w_3}{w_3(z_3-f)}e^{-\frac{\rho_3^{2}}{w_{3}^{2}(z_3-f)}}e^{\imath[k\frac{\rho_3^2}{2R(z_3-f)}-\eta(z_3-f)]}
\label{eq.A3b}
\ee 
where   $w_{3}(z)=w_{3}\sqrt{1+(\frac{z}{z_{R}})^2}$, $R(z)=z+\frac{z_{R}^2}{z}$,  $\eta(z)=\arctan(\frac{z}{z_{R}})$, and $z_{R}=\pi w_{3}^2/\lambda$.
The prefactor of $-i$ is the Gouy phase due to propagation from the front to back focal planes of the lens. 
 The second term is more complicated and with some manipulation can be put in the form
\bea
A_{3a}(\rho_3,s)&=&A_0(0) \frac{4h }{s}e^{\imath \frac{\rho_3^2}{w_0 w_3 s}}
\nn\\
&&\times \int_0^a d\rho_1\, \rho_1  
e^{\left(-h+ \frac{\imath}{w_0 w_3 s}\right)\rho_1^2}J_0\left(\frac{2\rho_1\rho_3}{w_0 w_3 s}\right).\nn
\eea
with $h= k^2 w_0^2/(4f^2-i 2fkw_0^2)=1/(w_3^2-i w_0 w_3)$. 
We have also introduced a normalized axial coordinate $s=z_3/f$, so that $s=1$ corresponds to the back focal plane. 
This last integral arises in the problem of diffraction from a finite circular aperture and can be expressed in terms of  Lommel functions of two variables (\cite{Whittaker1927}, p. 540)
\bse\bea
U_1(u,v)&=&u\int_0^1dt\, t J_0(v t)\cos\left[\frac{u}{2}(1-t^2) \right]\\
U_2(u,v)&=&u\int_0^1dt\, t J_0(v t)\sin\left[\frac{u}{2}(1-t^2) \right].
\eea
\label{eq.Lommel1}\ese
Using these definitions the field can be written as 
$$
A_{3a}(\rho_3,s)=
A_0(0)\frac{4 h a^2 e^{-\imath u/2}}{ s u}e^{\imath \frac{\rho_3^2}{w_0 w_3 s}}
\left[U_1(u,v)+i U_2(u,v) \right]
$$
with
$$
u = -2a^2\frac{w_3^2-w_0^2(s-1)+iw_0w_3 s}
{w_0 w_3 s(w_0^2+w_3^2)},~~~ v=\frac{2 a\rho_3}{w_0w_3s}.
$$
The total field in the focal region is thus
\be
A_3(\rho_3,s) = A_0(0)\frac{w_0}{w_3} A_{3b}(\rho_3,f s)+A_{3a}(\rho_3,s).
\label{eq.BBTfield}
\ee
The field and intensity distributions in the BBT can now be found from numerical evaluation of  the Lommel 
functions. We have also checked that direct numerical solution of the Fresnel propagation problem gives results which agree very closely with (\ref{eq.BBTfield}).
 
\begin{figure}[!t]
\begin{center}
\includegraphics[width=.4\textwidth]
{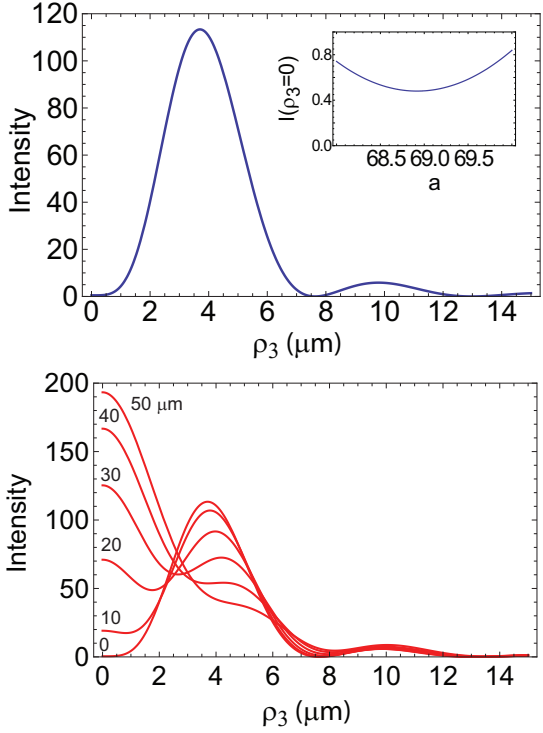}
\caption{(color online) BBT intensity profiles for $w_0=82.76~\mu\rm m$, $w_3=3~\mu\rm m$, $a=69~\mu\rm m$ and $s=1$ (upper plot). These parameters correspond to $\lambda=0.78~\mu\rm m$ and $f=1~\rm mm$. The plotted intensity is normalized to the peak of the input Gaussian with waist $w_0$.  The inset shows that the on-axis intensity does not vanish for any $a$. In the lower plot the transverse profiles are shown at axial displacements $z_3-f$ up to $50 ~\mu\rm m$ from the focal plane. }
\label{fig.HOEBoBprofile}
\end{center}
\end{figure}

\begin{figure}[!t]
\begin{center}
\vspace{-.1cm}
\includegraphics[width=.4\textwidth]
{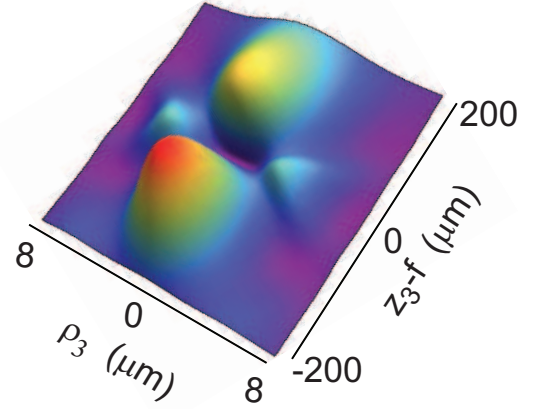}
\caption{(color online) Three dimensional intensity distribution about the focus. Same parameters as in Fig. \ref{fig.HOEBoBprofile}. }
\label{fig.hoebob3d}
\end{center}
\end{figure}

The intensity is plotted in Fig. \ref{fig.HOEBoBprofile} for some representative parameters. Note that the on-axis intensity does not go to zero for any value of $a$ due to the small diffractive phase shift which prevents perfect field cancellation. This could be corrected for by using a phase shift that is slightly different from $\pi$.  As the axial plane is moved away from the focus the on-axis intensity grows. We see that the minimum of the potential is about $2/3$ of the  radial peak in the focal plane. Figure \ref{fig.hoebob3d} shows that the intensity profile is not symmetric about the focal plane as $z$ is varied, with the peak on-axis intensity occurring in front of the bottle center.

Let us calculate  the intensity in  the focal plane  $z_3=f$ or $s=1$. We find
\bea
I(\rho_3,s=1)&=& \frac{\epsilon_0 c}{2}  \left|-iA_0(0)\frac{w_0}{w_3} e^{-\rho_3^2/w_3^2}+A_{3a}(\rho_3,1)\right|^2.\nn
\eea 
The on-axis intensity at $\rho_3=0$ should be as small as possible for a bottle beam. On axis 
\bea
A_{3a}(0,1)&=&
-i A_0(0)\frac{4 h a^2 }{ u}
\left(1-e^{-\imath u/2} \right), \nn
\eea
where we have used  
$U_1(u,0)=\sin(u/2), U_2(u,0)=2\sin^2(u/4).$ 
The intensity is thus 
$$
I(0,1) = \frac{\epsilon_0 c |A_0(0)|^2 w_0^2}{2w_3^2} \left|1- 2e^{\imath a^2 h w_3/w_0}\right|^2.
$$
Taking the derivative with respect to the phase plate radius $a$ and solving the resulting equation in the limit of $w_3\ll w_0$ gives the approximate condition for minimum intensity of  
$$
a=\sqrt{\ln2}\, w_0.
$$
This agrees with the condition found in \cite{Ozeri1999} which arises from simply setting the beam power transmitted through the $\pi$ plate equal to that in the outer annulus. Using the parameters of Fig. \ref{fig.HOEBoBprofile} ($w_0=82.76~\mu\rm m$) we get $a=68.9~\mu\rm m $ which 
agrees well with the value of $69.0~\mu\rm m$ used for the  full numerical solution shown in the figure. 

\subsection{Trap depth}

It is useful to have a simple expression for the trap depth created by the bottle beam. This is set by the intensity at the saddle region between the radial and axial intensity maxima.  A TEM$_{00}$ Gaussian beam has an intensity profile 
$$
I(r) = \frac{2P}{\pi w^2}e^{-2 r^2/w^2}
$$
where $P$ is the beam power. The bottle beam trap has a quartic radial profile in the back focal plane $(s=1)$. We can approximately model it as 
$$
I_B(r) = \frac{4P}{\pi w_B^6} r^4 e^{-2 r^2/w_B^2}.
$$
The prefactor has been chosen such that $\int_0^\infty dr\, 2\pi r I_B(r)=P.$ This beam has a ring structure with a maximum at $r=w_B$. At the maximum the intensity is 
$$
I_{B,\rm max}=I_B(w_B) = \frac{4P}{\pi w_B^2} e^{-2}.
$$
Assume the intensity at the saddle position is $q$ times the peak radial intensity. Then the effective trap intensity is 
$$
I_{\rm trap} \simeq q\frac{4P}{\pi w_B^2} \frac{1}{e^{2}}.
$$

Cs atoms trapped by 780 nm light have a  scalar polarizability of $\alpha_0^{\rm cgs} = -244. \times 10^{-24}~\rm cm^3$. In SI units we have $\alpha_0^{\rm SI}=4\pi\epsilon_0 10^{-6} \alpha_0^{\rm cgs}$. The light induced potential is $U=-\frac{1}{4}\alpha |\mcE|^2$ where the intensity is $I=\frac{\epsilon_0 c}{2}|\mcE|^2$ so, in SI units, 
$$
U_{\rm trap} = -\frac{1}{4}\alpha_0 \frac{2 I_{\rm trap}}{\epsilon_0 c} = -q\frac{\alpha_0^{\rm SI}}{\pi\epsilon_0 c}\frac{2P}{e^2 w_B^2}.
$$

The corresponding result for a TEM$_{00}$ Gaussian beam is
$$
U_{00} =-\frac{\alpha_0^{\rm SI}}{\pi \epsilon_0 c}\frac{P}{w^2}.  
$$
We see that the bottle beam trap with the same power has a lower trapping potential with the ratio $(2q/e^2)(w_0/w_B)^2$. 
Taking $P=100 ~\rm mW$, $q=0.6$, and $w_B=8~\mu\rm m$ gives $U_{\rm trap}/k_B= 40.~\mu\rm K$.

\section{Optical characterization}

\label{sec.doemeasurements}

\begin{figure}[!t]
\includegraphics[width=.45\textwidth]
{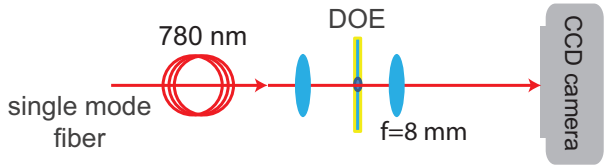}
\caption{\label{fig:DOEtestSetup} (color online) Setup for DOE testing.
A 780 nm beam from a single mode optical fiber is focused to a waist of 90 $\mu$m at the DOE. 
The BBT  created by the DOE is then imaged onto a CCD camera using an aspherical lens (Thorlabs C240TME-B) with $f=8 ~\rm mm$ and NA=0.5.  
The image magnification was measured with a USAF test chart to be 22.5.
  }
\end{figure}

\begin{figure}[!t]
\includegraphics[width=.5\textwidth]{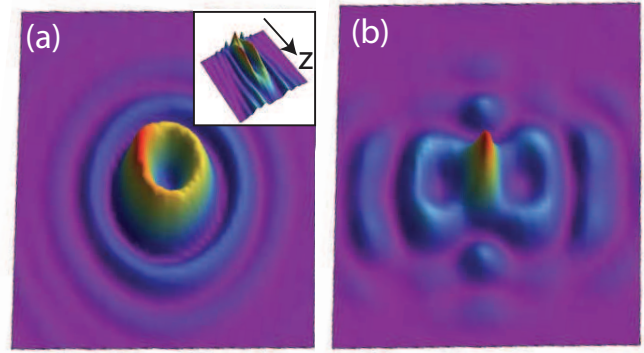}
\caption{\label{fig:BBTpics} (color online) Images of intensity distributions at the focal plane for single(left) and double(right) BBT.
The inset in a) shows the reconstructed 3D intensity distribution which was found by imaging different axial planes onto the CCD camera.
For these pictures we used a DOE with f=1 mm and $a= 57.18 ~\mu$m, the images are $35\times 35~\mu\rm m$.  }
\end{figure}

\begin{figure}[!t]
\includegraphics[width=70 mm]{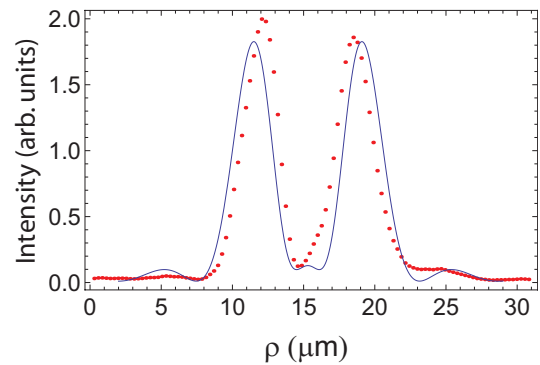}\rule{5mm}{0pt}
\caption{\label{fig:Intvsr} (color online) Intensity profile of the BBT recorded on a CCD camera.
The measured distribution (red points) is scaled by the
magnification of the imaging telescopes and compared to the theoretical calculations (solid blue line).
We used a DOE with f=1 mm and $a=57.18 ~\mu\rm m$. 
 }
\end{figure}

In this section we present optical measurements of the $f=1~\rm mm$ DOE. The test setup is shown in Fig.\ref{fig:DOEtestSetup}.
Pictures of single and double BBT intensity distributions (both with a focal distance of 1 mm, and predicted BBT trapped radius of
 3 $\mu$m) are shown in Fig.\ref{fig:BBTpics} (a) and (b) correspondingly.
Low intensity regions surrounded by high intensity barriers
are clearly visible. We verified the presence of a bottle structure in the longitudinal direction, as is shown in the inset.
The double BBT structure is formed by a linear grating on the DOE which is part of the fabricated DOE. 
We compare the observed intensity profile with  calculations in  Fig.\ref{fig:Intvsr}.
We observe nearly zero intensity at the center of the BBT, and the 3 $\mu$m radius of the peak of the ring  is in  good agreement with the calculated profile.

\begin{figure}[!t]
\includegraphics[width=70 mm]{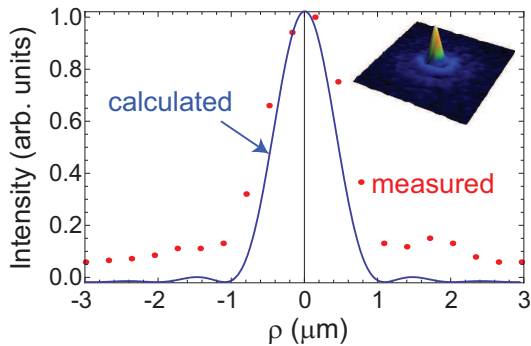}\rule{5mm}{0pt}
\caption{\label{fig:850nm} (color online) Focal spot created by 850 nm annular lens. The focal spot is imaged onto a CCD camera with magnification $31\times$ using a telescope with $f=8.0~\rm  mm$ aspheric lens and $f=250~\rm mm$ singlet lens. 
A cut through the center of the imaged beam (red points) is fitted by  Eq. (\ref{eq.annular}) (solid blue line). The inset shows the 2D intensity distribution in the focal plane. 
}
\end{figure}

We studied the focusing resolution of the annular DOE collection lens for 850 nm light by projecting a wide collimated beam  through the back side of the DOE.
The  collimated 850 nm beam passing through the Fresnel lens is focused down to a sub-micron waist and then imaged onto a CCD camera.
In  Fig.\ref{fig:850nm} we present the images of the 850 nm focused beam and a theoretical fit using
\be
I(\rho) =-\frac{I_0}{(1-\epsilon^2)^2}\left[\frac{2 J_1(x)}{x}-\frac{2 \epsilon J_1(\epsilon x)}{x}\right]^2.
\label{eq.annular}
\ee
Here $\epsilon=a/b=0.057$, $ J_1$ is the first order Bessel function, $I_0$ is the intensity of the incident beam 
and $x=\pi \rho/ \lambda$. The 850 nm beam is seen to be tightly focused with a beam waist ($1/e^2$ intensity radius) of about $1~\mu\rm m$.   The calculated profile using (\ref{eq.annular}) has a waist of $0.7~\mu\rm m$ which is in qualitative agreement with the observed focusing. The central focused region sits atop a small but broad background which may be due to defocus from the inner lens which is designed for 780 nm and is therefore aberrated at 850 nm.

\section{Atom trapping demonstration}

\label{sec.atoms}

\begin{figure}[!t]
\includegraphics[width=8.5cm]{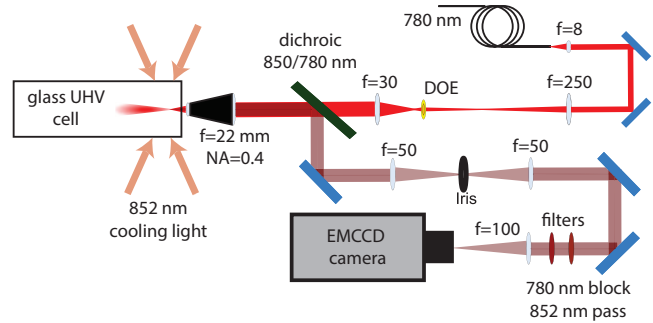}
\caption{\label{fig:Setup} (color online) Experimental setup for atom trapping in the BBT. 
Magneto-optical trap (MOT) of $^{133}$Cs is located in a glass UHV cell. The BBT is imaged into 
the vacuum chamber using a multielement lens with NA=0.4 which is diffraction limited at both 780 and 850 nm.
All other lenses have their focal lengths indicated in mm. The optics shown gave a BBT with radius to the maximum of the  intensity $w_B=3.1 ~\mu$m, giving a trap depth of about $400 ~\mu\rm K$ using 150 mW of 780 nm light. Data was also taken with other relay optics giving a trap with $w_B=5.8~\mu\rm m$  giving a  trap depth of about $230 ~\mu$K using 300 mW of 780 nm trapping light. 
The fluorescence from the trapped atoms is imaged onto an electron multiplying charge coupled device camera (EMCCD). }
\end{figure}

We used a  $^{133}$Cs magneto-optical trap (MOT) as a source of cold atoms for our BBT as shown in Fig. \ref{fig:Setup}.
The Cs MOT is operated in a standard configuration with three retro-reflected beams and loaded from a background cesium vapor.
Typically we collect 4$\times 10^{5}$  atoms in the MOT that are further cooled using polarization
gradient cooling (PGC).
The above described BBT is imaged into the vacuum chamber using a pair of relay lenses.  Imaging the BBT  into the vacuum chamber
rather than placing the DOE inside has a few practical  advantages. First, this allows additional flexibility of magnifying or demagnifying the BBT, and changing the DOE.
Second,  the DOE and the MOT can not be co-located since, the DOE would block some of the MOT beams. Hence
one would have to transport cold atoms from the MOT to the trapping region close to the DOE.
Since the BBT has a repulsive barrier around the trapping region, cold atoms outside the trap with low kinetic
energy will not enter and be trapped, hence one has to switch on the BBT during the PGC stage when the atoms are already present and cooled in the trap region.

 The atom loading sequence is  as follows.
First  Cs atoms are loaded into the MOT from the background vapor for 2 s.  Each MOT beam
has an intensity of about 2 mW/cm$^{2}$ and is 10 MHz red-detuned from the $F = 4 \rightarrow F' =5 $ cycling
transition at 852nm. 
 The atoms are further cooled using PGC for 3 ms  to reach temperatures of about 13 $\mu$K. During this phase the MOT magnetic field is switched off, the detuning of the MOT beams
is increased to  40 MHz, and their intensity is decreased by a factor of two. The BBT is turned on, atoms are further cooled in the BBT using PGC for another 3 ms.
 The MOT light is then extinguished for 45 ms, allowing untrapped MOT atoms to fall down away from the BBT region.
Finally, the MOT beams are turned on again for readout. Fluorescence light from  the trapped atomic cloud is imaged onto the CCD camera.

\begin{figure}[!t]
\includegraphics[width=9cm]{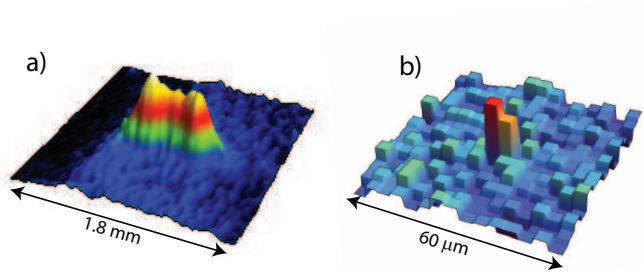}
\caption{\label{fig:TrappedAtoms} (color online) a) Side view of the fluorescence image of the atomic cloud 
trapped in a BBT with $w_B=5.8~\mu\rm m$. 
The image is the result of averaging 20 exposures  and substraction of the background, i.e. the image without any trapped atoms.
Roughly about 50-100 Cs atoms were trapped in the BBT potential. 
b) Axial view of about 10 atoms trapped in a   BBT with $w_B=3.1~\mu\rm m$. Each pixel is $3.9\times 3.9 ~\mu\rm m^2 $. }
\end{figure}

Fluorescence images of trapped atoms are  shown in Fig. \ref{fig:TrappedAtoms}.
Images were taken from the side, showing an atomic cloud elongated along the axis of the trap, and in backscatter. Estimates of the number of trapped atoms are based on the integrated fluorescence signal on the camera together with the theoretical single atom scattering rate. With the smaller BBT in Fig.  \ref{fig:TrappedAtoms}b) we observe small clouds of approximately 10 atoms localized to a few microns in the radial direction.   
We estimate the peak density of the atomic cloud as approximately  4$\times10^{10}$ cm$^{-3}$.

\begin{figure}[!t]
\includegraphics[width=65 mm]{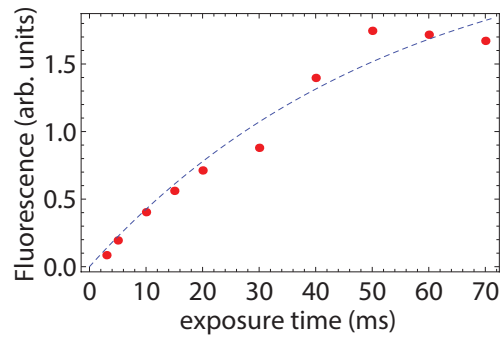}
\caption{\label{fig:Fluorescencevstime} (color online) Integrated fluorescence of the trapped atoms versus the exposure time.
Fluorescence images as in Fig.\ref{fig:TrappedAtoms}a) were recorded for various exposure time.
 Count numbers were summed along the horizontal direction and the obtained distribution was fitted by a Gaussian. The integrated fluorescence was
 defined as the area under the Gaussian fit. }
\end{figure}

We measured the integrated fluorescence of the trapped atoms in the larger BBT versus the exposure time as shown in Fig. \ref{fig:Fluorescencevstime}. 
If the number of  trapped
atoms  stayed constant during the exposure, one would have  expected a linear dependence of the integrated fluorescence vs. exposure time.
There are a number of reasons why atoms escape from the trap, such as collisions with background vapor, light assisted collisions,
and  motional heating caused by scattered MOT light. Although these loss mechanisms lead to different dependencies of the atom number versus exposure time our data is not sufficient to  differentiate between them. Assuming that the atom number drops exponentially we obtain a decay time of 60 ms.
We expected the vacuum limited life-time to be around 370 ms based on \cite{Bjorkholm1988}, which is roughly consistent with our measured MOT loading time
of 450 ms.
Although the equilibrium energy of the trapped atoms is well bellow the trap depth, the energy fluctuates due to interaction with the MOT light. 
Occasionally the atom energy exceeds the trap depth, and then the atom has a  chance to escape the trap. This causes  atomic loss with a time constant that  is consistent with 
the measured decay time of 60 ms. 
We anticipate that substantially longer lifetimes similar to the more than 
$5 ~\rm s$ observed in crossed vortex beam traps\cite{Li2012} will be reached  in a deeper BBT and with better vacuum conditions.

\section{Conclusions}

\label{sec.end}

We have studied both theoretically and experimentally BBTs created by a DOE. We have demonstrated trapping of laser cooled  Cs
atoms in the BBT potential. The BBTs produced by DOEs are promising for creation of 
complex trapping potentials using a single optical element. Such an approach has advantages for miniaturization
and integration of the optical setup. The BBT potentials can be used for trapping both ground and Rydberg state atoms\cite{SZhang2011}, which is crucial for quantum computing and precision spectroscopy of  Rydberg levels.

\begin{acknowledgments}

This  work was supported by the Sandia LDRD program in adiabatic quantum computation.
Sandia is a multiprogram laboratory operated by Sandia Corporation, a Lockheed Martin Company, for the United States Department of Energy's National Nuclear Security Administration under contract DE-AC04-94AL85000.

\end{acknowledgments}
  

\begin{thebibliography}{21}%
\makeatletter
\providecommand \@ifxundefined [1]{%
 \@ifx{#1\undefined}
}%
\providecommand \@ifnum [1]{%
 \ifnum #1\expandafter \@firstoftwo
 \else \expandafter \@secondoftwo
 \fi
}%
\providecommand \@ifx [1]{%
 \ifx #1\expandafter \@firstoftwo
 \else \expandafter \@secondoftwo
 \fi
}%
\providecommand \natexlab [1]{#1}%
\providecommand \enquote  [1]{``#1''}%
\providecommand \bibnamefont  [1]{#1}%
\providecommand \bibfnamefont [1]{#1}%
\providecommand \citenamefont [1]{#1}%
\providecommand \href@noop [0]{\@secondoftwo}%
\providecommand \href [0]{\begingroup \@sanitize@url \@href}%
\providecommand \@href[1]{\@@startlink{#1}\@@href}%
\providecommand \@@href[1]{\endgroup#1\@@endlink}%
\providecommand \@sanitize@url [0]{\catcode `\\12\catcode `\$12\catcode
  `\&12\catcode `\#12\catcode `\^12\catcode `\_12\catcode `\%12\relax}%
\providecommand \@@startlink[1]{}%
\providecommand \@@endlink[0]{}%
\providecommand \url  [0]{\begingroup\@sanitize@url \@url }%
\providecommand \@url [1]{\endgroup\@href {#1}{\urlprefix }}%
\providecommand \urlprefix  [0]{URL }%
\providecommand \Eprint [0]{\href }%
\providecommand \doibase [0]{http://dx.doi.org/}%
\providecommand \selectlanguage [0]{\@gobble}%
\providecommand \bibinfo  [0]{\@secondoftwo}%
\providecommand \bibfield  [0]{\@secondoftwo}%
\providecommand \translation [1]{[#1]}%
\providecommand \BibitemOpen [0]{}%
\providecommand \bibitemStop [0]{}%
\providecommand \bibitemNoStop [0]{.\EOS\space}%
\providecommand \EOS [0]{\spacefactor3000\relax}%
\providecommand \BibitemShut  [1]{\csname bibitem#1\endcsname}%
\let\auto@bib@innerbib\@empty
\bibitem [{\citenamefont {Meschede}\ and\ \citenamefont
  {Rauschenbeutel}(2006)}]{Meschede2006}%
  \BibitemOpen
  \bibfield  {author} {\bibinfo {author} {\bibfnamefont {D.}~\bibnamefont
  {Meschede}}\ and\ \bibinfo {author} {\bibfnamefont {A.}~\bibnamefont
  {Rauschenbeutel}},\ }\href@noop {} {\bibfield  {journal} {\bibinfo  {journal}
  {Adv. At. Mol. Opt. Phys.}\ }\textbf {\bibinfo {volume} {53}},\ \bibinfo
  {pages} {75} (\bibinfo {year} {2006})}\BibitemShut {NoStop}%
\bibitem [{\citenamefont {Saffman}\ \emph {et~al.}(2010)\citenamefont
  {Saffman}, \citenamefont {Walker},\ and\ \citenamefont
  {M\o{}lmer}}]{Saffman2010}%
  \BibitemOpen
  \bibfield  {author} {\bibinfo {author} {\bibfnamefont {M.}~\bibnamefont
  {Saffman}}, \bibinfo {author} {\bibfnamefont {T.~G.}\ \bibnamefont {Walker}},
  \ and\ \bibinfo {author} {\bibfnamefont {K.}~\bibnamefont {M\o{}lmer}},\
  }\href@noop {} {\bibfield  {journal} {\bibinfo  {journal} {Rev. Mod. Phys.}\
  }\textbf {\bibinfo {volume} {82}},\ \bibinfo {pages} {2313} (\bibinfo {year}
  {2010})}\BibitemShut {NoStop}%
\bibitem [{\citenamefont {Negretti}\ \emph {et~al.}(2011)\citenamefont
  {Negretti}, \citenamefont {Treutlein},\ and\ \citenamefont
  {Calarco}}]{Negretti2011}%
  \BibitemOpen
  \bibfield  {author} {\bibinfo {author} {\bibfnamefont {A.}~\bibnamefont
  {Negretti}}, \bibinfo {author} {\bibfnamefont {P.}~\bibnamefont {Treutlein}},
  \ and\ \bibinfo {author} {\bibfnamefont {T.}~\bibnamefont {Calarco}},\
  }\href@noop {} {\bibfield  {journal} {\bibinfo  {journal} {Quantum Inf.
  Process.}\ }\textbf {\bibinfo {volume} {10}},\ \bibinfo {pages} {721}
  (\bibinfo {year} {2011})}\BibitemShut {NoStop}%
\bibitem [{\citenamefont {Schlosser}\ \emph {et~al.}(2011)\citenamefont
  {Schlosser}, \citenamefont {Tichelmann}, \citenamefont {Kruse},\ and\
  \citenamefont {Birkl}}]{Schlosser2011}%
  \BibitemOpen
  \bibfield  {author} {\bibinfo {author} {\bibfnamefont {M.}~\bibnamefont
  {Schlosser}}, \bibinfo {author} {\bibfnamefont {S.}~\bibnamefont
  {Tichelmann}}, \bibinfo {author} {\bibfnamefont {J.}~\bibnamefont {Kruse}}, \
  and\ \bibinfo {author} {\bibfnamefont {G.}~\bibnamefont {Birkl}},\
  }\href@noop {} {\bibfield  {journal} {\bibinfo  {journal} {Quantum Inf.
  Process.}\ }\textbf {\bibinfo {volume} {10}},\ \bibinfo {pages} {907}
  (\bibinfo {year} {2011})}\BibitemShut {NoStop}%
\bibitem [{\citenamefont {Streed}\ \emph {et~al.}(2011)\citenamefont {Streed},
  \citenamefont {Norton}, \citenamefont {Jechow}, \citenamefont {Weinhold},\
  and\ \citenamefont {Kielpinski}}]{Streed2011}%
  \BibitemOpen
  \bibfield  {author} {\bibinfo {author} {\bibfnamefont {E.~W.}\ \bibnamefont
  {Streed}}, \bibinfo {author} {\bibfnamefont {B.~G.}\ \bibnamefont {Norton}},
  \bibinfo {author} {\bibfnamefont {A.}~\bibnamefont {Jechow}}, \bibinfo
  {author} {\bibfnamefont {T.~J.}\ \bibnamefont {Weinhold}}, \ and\ \bibinfo
  {author} {\bibfnamefont {D.}~\bibnamefont {Kielpinski}},\ }\href@noop {}
  {\bibfield  {journal} {\bibinfo  {journal} {Phys. Rev. Lett.}\ }\textbf
  {\bibinfo {volume} {106}},\ \bibinfo {pages} {010502} (\bibinfo {year}
  {2011})}\BibitemShut {NoStop}%
\bibitem [{\citenamefont {Kim}\ \emph {et~al.}(2011)\citenamefont {Kim},
  \citenamefont {Maunz},\ and\ \citenamefont {Kim}}]{Kim2011}%
  \BibitemOpen
  \bibfield  {author} {\bibinfo {author} {\bibfnamefont {T.}~\bibnamefont
  {Kim}}, \bibinfo {author} {\bibfnamefont {P.}~\bibnamefont {Maunz}}, \ and\
  \bibinfo {author} {\bibfnamefont {J.}~\bibnamefont {Kim}},\ }\href@noop {}
  {\bibfield  {journal} {\bibinfo  {journal} {Phys. Rev. A}\ }\textbf {\bibinfo
  {volume} {84}},\ \bibinfo {pages} {063423} (\bibinfo {year}
  {2011})}\BibitemShut {NoStop}%
\bibitem [{\citenamefont {Brady}\ \emph {et~al.}(2011)\citenamefont {Brady},
  \citenamefont {Ellis}, \citenamefont {Moehring}, \citenamefont {Stick},
  \citenamefont {Highstrete}, \citenamefont {Fortier}, \citenamefont {Blain},
  \citenamefont {Haltli}, \citenamefont {Cruz-Cabrera}, \citenamefont {Briggs},
  \citenamefont {Wendt}, \citenamefont {Carter}, \citenamefont {Samora},\ and\
  \citenamefont {Kemme}}]{Brady2011}%
  \BibitemOpen
  \bibfield  {author} {\bibinfo {author} {\bibfnamefont {G.}~\bibnamefont
  {Brady}}, \bibinfo {author} {\bibfnamefont {A.}~\bibnamefont {Ellis}},
  \bibinfo {author} {\bibfnamefont {D.}~\bibnamefont {Moehring}}, \bibinfo
  {author} {\bibfnamefont {D.}~\bibnamefont {Stick}}, \bibinfo {author}
  {\bibfnamefont {C.}~\bibnamefont {Highstrete}}, \bibinfo {author}
  {\bibfnamefont {K.}~\bibnamefont {Fortier}}, \bibinfo {author} {\bibfnamefont
  {M.}~\bibnamefont {Blain}}, \bibinfo {author} {\bibfnamefont
  {R.}~\bibnamefont {Haltli}}, \bibinfo {author} {\bibfnamefont
  {A.}~\bibnamefont {Cruz-Cabrera}}, \bibinfo {author} {\bibfnamefont
  {R.}~\bibnamefont {Briggs}}, \bibinfo {author} {\bibfnamefont
  {J.}~\bibnamefont {Wendt}}, \bibinfo {author} {\bibfnamefont
  {T.}~\bibnamefont {Carter}}, \bibinfo {author} {\bibfnamefont
  {S.}~\bibnamefont {Samora}}, \ and\ \bibinfo {author} {\bibfnamefont
  {S.}~\bibnamefont {Kemme}},\ }\href@noop {} {\bibfield  {journal} {\bibinfo
  {journal} {Applied Physics B}\ }\textbf {\bibinfo {volume} {103}},\ \bibinfo
  {pages} {801} (\bibinfo {year} {2011})}\BibitemShut {NoStop}%
\bibitem [{\citenamefont {Xu}\ \emph {et~al.}(2010)\citenamefont {Xu},
  \citenamefont {He}, \citenamefont {Wang},\ and\ \citenamefont
  {Zhan}}]{Xu2010}%
  \BibitemOpen
  \bibfield  {author} {\bibinfo {author} {\bibfnamefont {P.}~\bibnamefont
  {Xu}}, \bibinfo {author} {\bibfnamefont {X.}~\bibnamefont {He}}, \bibinfo
  {author} {\bibfnamefont {J.}~\bibnamefont {Wang}}, \ and\ \bibinfo {author}
  {\bibfnamefont {M.}~\bibnamefont {Zhan}},\ }\href@noop {} {\bibfield
  {journal} {\bibinfo  {journal} {Opt. Lett.}\ }\textbf {\bibinfo {volume}
  {35}},\ \bibinfo {pages} {2164} (\bibinfo {year} {2010})}\BibitemShut
  {NoStop}%
\bibitem [{\citenamefont {Li}\ \emph {et~al.}(2012)\citenamefont {Li},
  \citenamefont {Zhang}, \citenamefont {Isenhower}, \citenamefont {Maller},\
  and\ \citenamefont {Saffman}}]{Li2012}%
  \BibitemOpen
  \bibfield  {author} {\bibinfo {author} {\bibfnamefont {G.}~\bibnamefont
  {Li}}, \bibinfo {author} {\bibfnamefont {S.}~\bibnamefont {Zhang}}, \bibinfo
  {author} {\bibfnamefont {L.}~\bibnamefont {Isenhower}}, \bibinfo {author}
  {\bibfnamefont {K.}~\bibnamefont {Maller}}, \ and\ \bibinfo {author}
  {\bibfnamefont {M.}~\bibnamefont {Saffman}},\ }\href@noop {} {\bibfield
  {journal} {\bibinfo  {journal} {Opt. Lett.}\ }\textbf {\bibinfo {volume}
  {37}},\ \bibinfo {pages} {851} (\bibinfo {year} {2012})}\BibitemShut
  {NoStop}%
\bibitem [{\citenamefont {Zhang}\ \emph {et~al.}(2011)\citenamefont {Zhang},
  \citenamefont {Robicheaux},\ and\ \citenamefont {Saffman}}]{SZhang2011}%
  \BibitemOpen
  \bibfield  {author} {\bibinfo {author} {\bibfnamefont {S.}~\bibnamefont
  {Zhang}}, \bibinfo {author} {\bibfnamefont {F.}~\bibnamefont {Robicheaux}}, \
  and\ \bibinfo {author} {\bibfnamefont {M.}~\bibnamefont {Saffman}},\
  }\href@noop {} {\bibfield  {journal} {\bibinfo  {journal} {Phys. Rev. A}\
  }\textbf {\bibinfo {volume} {84}},\ \bibinfo {pages} {043408} (\bibinfo
  {year} {2011})}\BibitemShut {NoStop}%
\bibitem [{\citenamefont {Wilk}\ \emph {et~al.}(2010)\citenamefont {Wilk},
  \citenamefont {Ga\"etan}, \citenamefont {Evellin}, \citenamefont {Wolters},
  \citenamefont {Miroshnychenko}, \citenamefont {Grangier},\ and\ \citenamefont
  {Browaeys}}]{Wilk2010}%
  \BibitemOpen
  \bibfield  {author} {\bibinfo {author} {\bibfnamefont {T.}~\bibnamefont
  {Wilk}}, \bibinfo {author} {\bibfnamefont {A.}~\bibnamefont {Ga\"etan}},
  \bibinfo {author} {\bibfnamefont {C.}~\bibnamefont {Evellin}}, \bibinfo
  {author} {\bibfnamefont {J.}~\bibnamefont {Wolters}}, \bibinfo {author}
  {\bibfnamefont {Y.}~\bibnamefont {Miroshnychenko}}, \bibinfo {author}
  {\bibfnamefont {P.}~\bibnamefont {Grangier}}, \ and\ \bibinfo {author}
  {\bibfnamefont {A.}~\bibnamefont {Browaeys}},\ }\href@noop {} {\bibfield
  {journal} {\bibinfo  {journal} {Phys. Rev. Lett.}\ }\textbf {\bibinfo
  {volume} {104}},\ \bibinfo {pages} {010502} (\bibinfo {year}
  {2010})}\BibitemShut {NoStop}%
\bibitem [{\citenamefont {Isenhower}\ \emph {et~al.}(2010)\citenamefont
  {Isenhower}, \citenamefont {Urban}, \citenamefont {Zhang}, \citenamefont
  {Gill}, \citenamefont {Henage}, \citenamefont {Johnson}, \citenamefont
  {Walker},\ and\ \citenamefont {Saffman}}]{Isenhower2010}%
  \BibitemOpen
  \bibfield  {author} {\bibinfo {author} {\bibfnamefont {L.}~\bibnamefont
  {Isenhower}}, \bibinfo {author} {\bibfnamefont {E.}~\bibnamefont {Urban}},
  \bibinfo {author} {\bibfnamefont {X.~L.}\ \bibnamefont {Zhang}}, \bibinfo
  {author} {\bibfnamefont {A.~T.}\ \bibnamefont {Gill}}, \bibinfo {author}
  {\bibfnamefont {T.}~\bibnamefont {Henage}}, \bibinfo {author} {\bibfnamefont
  {T.~A.}\ \bibnamefont {Johnson}}, \bibinfo {author} {\bibfnamefont {T.~G.}\
  \bibnamefont {Walker}}, \ and\ \bibinfo {author} {\bibfnamefont
  {M.}~\bibnamefont {Saffman}},\ }\href@noop {} {\bibfield  {journal} {\bibinfo
   {journal} {Phys. Rev. Lett.}\ }\textbf {\bibinfo {volume} {104}},\ \bibinfo
  {pages} {010503} (\bibinfo {year} {2010})}\BibitemShut {NoStop}%
\bibitem [{\citenamefont {Zhang}\ \emph {et~al.}(2010)\citenamefont {Zhang},
  \citenamefont {Isenhower}, \citenamefont {Gill}, \citenamefont {Walker},\
  and\ \citenamefont {Saffman}}]{Zhang2010}%
  \BibitemOpen
  \bibfield  {author} {\bibinfo {author} {\bibfnamefont {X.~L.}\ \bibnamefont
  {Zhang}}, \bibinfo {author} {\bibfnamefont {L.}~\bibnamefont {Isenhower}},
  \bibinfo {author} {\bibfnamefont {A.~T.}\ \bibnamefont {Gill}}, \bibinfo
  {author} {\bibfnamefont {T.~G.}\ \bibnamefont {Walker}}, \ and\ \bibinfo
  {author} {\bibfnamefont {M.}~\bibnamefont {Saffman}},\ }\href@noop {}
  {\bibfield  {journal} {\bibinfo  {journal} {Phys. Rev. A}\ }\textbf {\bibinfo
  {volume} {82}},\ \bibinfo {pages} {030306(R)} (\bibinfo {year}
  {2010})}\BibitemShut {NoStop}%
\bibitem [{\citenamefont {Keating}\ \emph {et~al.}(2013)\citenamefont
  {Keating}, \citenamefont {Goyal}, \citenamefont {Jau}, \citenamefont
  {Biedermann}, \citenamefont {Landahl},\ and\ \citenamefont
  {Deutsch}}]{Keating2013}%
  \BibitemOpen
  \bibfield  {author} {\bibinfo {author} {\bibfnamefont {T.}~\bibnamefont
  {Keating}}, \bibinfo {author} {\bibfnamefont {K.}~\bibnamefont {Goyal}},
  \bibinfo {author} {\bibfnamefont {Y.-Y.}\ \bibnamefont {Jau}}, \bibinfo
  {author} {\bibfnamefont {G.~W.}\ \bibnamefont {Biedermann}}, \bibinfo
  {author} {\bibfnamefont {A.}~\bibnamefont {Landahl}}, \ and\ \bibinfo
  {author} {\bibfnamefont {I.~H.}\ \bibnamefont {Deutsch}},\ }\href@noop {}
  {\bibfield  {journal} {\bibinfo  {journal} {Phys. Rev. A}\ }\textbf {\bibinfo
  {volume} {87}},\ \bibinfo {pages} {052314} (\bibinfo {year}
  {2013})}\BibitemShut {NoStop}%
\bibitem [{\citenamefont {Carr}\ and\ \citenamefont
  {Saffman}(2013)}]{Carr2013}%
  \BibitemOpen
  \bibfield  {author} {\bibinfo {author} {\bibfnamefont {A.}~\bibnamefont
  {Carr}}\ and\ \bibinfo {author} {\bibfnamefont {M.}~\bibnamefont {Saffman}},\
  }\href@noop {} {\  (\bibinfo {year} {2013})},\ \bibinfo {note}
  {arXiv:1304.6374}\BibitemShut {NoStop}%
\bibitem [{\citenamefont {Chaloupka}\ \emph {et~al.}(1997)\citenamefont
  {Chaloupka}, \citenamefont {Fisher}, \citenamefont {Kessler},\ and\
  \citenamefont {Meyerhofer}}]{Chaloupka1997}%
  \BibitemOpen
  \bibfield  {author} {\bibinfo {author} {\bibfnamefont {J.~L.}\ \bibnamefont
  {Chaloupka}}, \bibinfo {author} {\bibfnamefont {Y.}~\bibnamefont {Fisher}},
  \bibinfo {author} {\bibfnamefont {T.~J.}\ \bibnamefont {Kessler}}, \ and\
  \bibinfo {author} {\bibfnamefont {D.~D.}\ \bibnamefont {Meyerhofer}},\
  }\href@noop {} {\bibfield  {journal} {\bibinfo  {journal} {Opt. Lett.}\
  }\textbf {\bibinfo {volume} {22}},\ \bibinfo {pages} {1021} (\bibinfo {year}
  {1997})}\BibitemShut {NoStop}%
\bibitem [{\citenamefont {Ozeri}\ \emph {et~al.}(1999)\citenamefont {Ozeri},
  \citenamefont {Khaykovich},\ and\ \citenamefont {Davidson}}]{Ozeri1999}%
  \BibitemOpen
  \bibfield  {author} {\bibinfo {author} {\bibfnamefont {R.}~\bibnamefont
  {Ozeri}}, \bibinfo {author} {\bibfnamefont {L.}~\bibnamefont {Khaykovich}}, \
  and\ \bibinfo {author} {\bibfnamefont {N.}~\bibnamefont {Davidson}},\
  }\href@noop {} {\bibfield  {journal} {\bibinfo  {journal} {Phys. Rev. A}\
  }\textbf {\bibinfo {volume} {59}},\ \bibinfo {pages} {R1750} (\bibinfo {year}
  {1999})},\ \bibinfo {note} {erratum: Phys. Rev. A {\bf 65}, 069903
  (2002)}\BibitemShut {NoStop}%
\bibitem [{\citenamefont {Ozeri}\ \emph {et~al.}(2002)\citenamefont {Ozeri},
  \citenamefont {Khaykovich},\ and\ \citenamefont {Davidson}}]{Ozeri2002}%
  \BibitemOpen
  \bibfield  {author} {\bibinfo {author} {\bibfnamefont {R.}~\bibnamefont
  {Ozeri}}, \bibinfo {author} {\bibfnamefont {L.}~\bibnamefont {Khaykovich}}, \
  and\ \bibinfo {author} {\bibfnamefont {N.}~\bibnamefont {Davidson}},\
  }\href@noop {} {\bibfield  {journal} {\bibinfo  {journal} {Phys. Rev. A}\
  }\textbf {\bibinfo {volume} {65}},\ \bibinfo {pages} {069903} (\bibinfo
  {year} {2002})}\BibitemShut {NoStop}%
\bibitem [{\citenamefont {Kemme}\ \emph {et~al.}(2012)\citenamefont {Kemme},
  \citenamefont {Brady}, \citenamefont {Ellis}, \citenamefont {Wendt},
  \citenamefont {Peters}, \citenamefont {Biedermann}, \citenamefont {Carter},
  \citenamefont {Samora}, \citenamefont {Isaacs}, \citenamefont {Ivanov},\ and\
  \citenamefont {Saffman}}]{Kemme2012}%
  \BibitemOpen
  \bibfield  {author} {\bibinfo {author} {\bibfnamefont {S.}~\bibnamefont
  {Kemme}}, \bibinfo {author} {\bibfnamefont {G.}~\bibnamefont {Brady}},
  \bibinfo {author} {\bibfnamefont {A.}~\bibnamefont {Ellis}}, \bibinfo
  {author} {\bibfnamefont {J.}~\bibnamefont {Wendt}}, \bibinfo {author}
  {\bibfnamefont {D.}~\bibnamefont {Peters}}, \bibinfo {author} {\bibfnamefont
  {G.}~\bibnamefont {Biedermann}}, \bibinfo {author} {\bibfnamefont
  {T.}~\bibnamefont {Carter}}, \bibinfo {author} {\bibfnamefont
  {S.}~\bibnamefont {Samora}}, \bibinfo {author} {\bibfnamefont
  {J.}~\bibnamefont {Isaacs}}, \bibinfo {author} {\bibfnamefont
  {V.}~\bibnamefont {Ivanov}}, \ and\ \bibinfo {author} {\bibfnamefont
  {M.}~\bibnamefont {Saffman}},\ }\href@noop {} {\bibfield  {journal} {\bibinfo
   {journal} {Proc. S.P.I.E.}\ }\textbf {\bibinfo {volume} {8249}},\ \bibinfo
  {pages} {82490E} (\bibinfo {year} {2012})}\BibitemShut {NoStop}%
\bibitem [{\citenamefont {Whittaker}\ and\ \citenamefont
  {Watson}(1927)}]{Whittaker1927}%
  \BibitemOpen
  \bibfield  {author} {\bibinfo {author} {\bibfnamefont {E.~T.}\ \bibnamefont
  {Whittaker}}\ and\ \bibinfo {author} {\bibfnamefont {G.~N.}\ \bibnamefont
  {Watson}},\ }\href@noop {} {\emph {\bibinfo {title} {A course of modern
  analysis, 4$^{th}$ Ed.}}}\ (\bibinfo  {publisher} {Cambridge University
  Press},\ \bibinfo {address} {Cambridge},\ \bibinfo {year} {1927})\BibitemShut
  {NoStop}%
\bibitem [{\citenamefont {Bjorkholm}(1988)}]{Bjorkholm1988}%
  \BibitemOpen
  \bibfield  {author} {\bibinfo {author} {\bibfnamefont {J.~E.}\ \bibnamefont
  {Bjorkholm}},\ }\href@noop {} {\bibfield  {journal} {\bibinfo  {journal}
  {Phys. Rev. A}\ }\textbf {\bibinfo {volume} {38}},\ \bibinfo {pages} {1599}
  (\bibinfo {year} {1988})}\BibitemShut {NoStop}%
\end{thebibliography}

%

\end{document}